\documentclass[%
 aip,
floatfix,amsmath,amssymb,
 reprint,%
]{revtex4-1}

\usepackage{graphicx}
\usepackage{dcolumn}
\usepackage{bm}
\usepackage{float}
\usepackage{epsfig}
\usepackage{epstopdf}

\usepackage[utf8]{inputenc}
\usepackage[T1]{fontenc}
\usepackage{mathptmx}

\newcommand{\lnsco}{La$_{1.48}$Nd$_{0.4}$Sr$_{0.12}$CuO$_{4}$}

\begin{document}

\preprint{AIP/123-QED}

\title{Charge order dynamics in underdoped La$\mathbf{_{1.6-\textit{x}}}$Nd$\mathbf{_{0.4}}$Sr$\mathbf{{_\textit{x}}}$CuO$\mathbf{_{4}}$
revealed by electric pulses} 

\author{Bal~K.~Pokharel}
\affiliation{National High Magnetic Field Laboratory, Tallahassee, Florida 32310, USA}
\affiliation{Department of Physics, Florida State University, Tallahassee, Florida 32306, USA}
\author{Yuxin~Wang}
\affiliation{National High Magnetic Field Laboratory, Tallahassee, Florida 32310, USA}
\affiliation{Department of Physics, Florida State University, Tallahassee, Florida 32306, USA}
\author{J.~Jaroszynski}
\affiliation{National High Magnetic Field Laboratory, Tallahassee, Florida 32310, USA}
\author{T.~Sasagawa}
\affiliation{Materials and Structures Laboratory, Tokyo Institute of Technology, Kanagawa 226-8503, Japan}
\author{Dragana~Popovi\'c}
\email{dragana@magnet.fsu.edu}
\affiliation{National High Magnetic Field Laboratory, Tallahassee, Florida 32310, USA}
\affiliation{Department of Physics, Florida State University, Tallahassee, Florida 32306, USA}

\date{\today}

\begin{abstract}
The dynamics of the charge-order domains has been investigated in \lnsco, a prototypical stripe-ordered cuprate, using pulsed current injection.  We first identify the regime in which nonthermal effects dominate over simple Joule heating, and then demonstrate that, for small enough perturbation, pulsed current injection allows access to nonthermally-induced resistive metastable states.  The results are consistent with pinning of the fluctuating charge order, with fluctuations being most pronounced at the charge-order onset temperature.  The nonequilibrium effects are revealed only when the transition is approached from the charge-ordered phase.  Our experiment establishes pulsed current injection as a viable and effective method for probing the charge-order dynamics in various other materials.
\end{abstract}

\maketitle

Charge density modulations or charge orders (COs) are observed in all families of hole-doped cuprate high-temperature superconductors,\cite{Comin2016} but their relevance for the unconventional properties of the normal state and superconductivity is still an open question.\cite{Fradkin2015,Huang2017,Zheng2017,Peng2018,Miao2021}  According to one broadly-considered scenario, fluctuations of the incipient CO could be favorable or even contribute to the pairing mechanism.\cite{Kivelson2003, MVojta2009}  Therefore, the existence of CO fluctuations and the nature of their dynamics are some of the key issues in the physics of cuprates.  Although detecting CO fluctuations has been a challenge because of the remarkable stability of the CO and its short-range nature, both believed to be due to the pinning by disorder, they have been reported recently in several cuprates over a wide range of doping.\cite{Torchinsky2013,Baity2018,Mitrano2019,Arpaia2019,Boschini2021}  However, relatively little is known about their dynamics.

We report a new technique to study the cuprate CO dynamics, in which we apply electrical pulses to drive the CO system out of equilibrium and then study its response using charge transport measurements.  Similar studies have been used previously to probe the dynamics of conventional charge-density-wave (CDW) systems, such as 1\textit{T}-TaS$_2$\cite{Vaskivskyi2016,Ma2019,Ma2020} and some organic conductors.\cite{Cohen1977,Kagawa2016}  More generally, electrical control and switching of resistive states by electric pulsing in strongly correlated materials is of great interest for the development of next generation of solid-state devices.\cite{Stoliar2013}  However, one of the main challenges has been to distinguish between the effects of Joule heating and nonthermal effects of the electric field.\cite{Giorgianni2019,Kalcheim2020}  We study \lnsco, in which CO is in the form of stripes,\cite{Tranquada1995} and demonstrate that, for small enough perturbation, pulsed current injection allows access to nonthermally-induced resistive metastable states.  The results are consistent with strong pinning of the fluctuating CO by disorder.  Our findings pave the way for similar studies in various stripe-ordered materials, such as other cuprates and nickelates.

\begin{figure}[!hbt]
    \centering
      \includegraphics[width=0.47\textwidth,clip=]{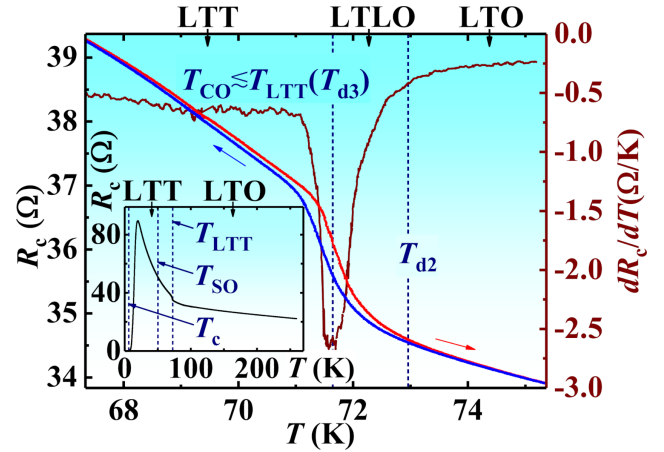}
    \caption{$R_{\mathrm{c}}$ vs $T$ across the structural and CO transition regions of La$_{1.48}$Nd$_{0.4}$Sr$_{0.12}$CuO$_4$ with a hysteresis loop (independent of sweep rates for 0.01-1~K/min); $T_{\mathrm{CO}}\lesssim T_{\mathrm{LTT}}\simeq 71.65$~K.  The features of $dR_{\mathrm{c}}/dT$ of the warming branch marked by the vertical dotted lines correspond to $T_{\mathrm{d}3}=T_{\textrm{LTT}}$ and $T_{\mathrm{d}2}$, the temperatures of the LTT-LTLO and LTLO-LTO transitions, respectively.\cite{Sakita1999}  Inset: $R_\mathrm{c}$ vs $T$.  $R_\mathrm{c}\rightarrow 0$ at $T_{\mathrm{c}}\simeq 3.5$~K; $T_\textrm{SO}\simeq 50$~K is the onset for spin stripe order.\cite{Tranquada1996,Zimmermann1998,Ichikawa2000}}  
    \label{fig:RT}
\end{figure}

La$_{2-x-y}$Sr$_x$(Nd,Eu)$_y$CuO$_4$ and La$_{2-x}$Ba$_x$CuO$_4$ compounds are cuprates that exhibit strongest CO correlations.  The striped CO is stabilized by the anisotropy within the CuO$_2$ planes that is present only in the low-temperature tetragonal (LTT) crystallographic phase.  Stripes are rotated by 90$^{\circ}$ from one CuO$_2$ layer to next\cite{Tranquada1995} and, just like in other cuprates, this CO is most pronounced for hole doping $x\approx 1/8$, corresponding to a minimum in bulk superconducting transition temperature $T_\mathrm{c}$.  In \lnsco, the onset of the apparent static CO occurs at $T=T_{\textrm{CO}}\lesssim T_{\textrm{LTT}}\simeq 71.65$~K, where $T_{\textrm{LTT}}$ is the transition temperature from the low-temperature orthorhombic (LTO) to LTT phase, with the transition consisting of a $45^{\circ}$ rotation of the tilting axis of the oxygen octahedra surrounding the Cu atoms.\cite{Hucker2012}  The LTO-LTT transition region is also characterized by the presence of an intermediate, low-temperature less-orthorhombic (LTLO) phase, in which the rotation of the octahedral tilt axis is not complete.  The structural transition region is manifested as a jump in the $c$-axis resistance $R_\mathrm{c}(T)$, accompanied by a thermal hysteresis (Fig.~\ref{fig:RT}),
which is attributed to the first-order nature of the structural transition.\cite{Nakamura1992} \lnsco\ is an ideal candidate for electrical pulse studies because evidence for metastable states, collective behavior, and criticality, signatures of fluctuating CO, were found\cite{Baity2018} in $R_\mathrm{c}(T)$ in the regime across the CO (and structural) transition following the application of a magnetic field ($H$) or a large change in $T$ as an external perturbation.  Surprisingly, those effects were revealed only when the transition region was approached from the CO phase.  The measurements were performed using a small, constant electric field $\sim 0.06$~V/cm.  Here, in contrast, we apply current pulses of different amplitude $I_\mathrm{p}$ and duration $\tau$ at a constant $T$, and measure the initial and final resistances with a low current $I=10~\mu$A before and after each pulse.

\begin{figure*}
    \centering
    \centering
                     \includegraphics[width=0.9\textwidth,clip=]{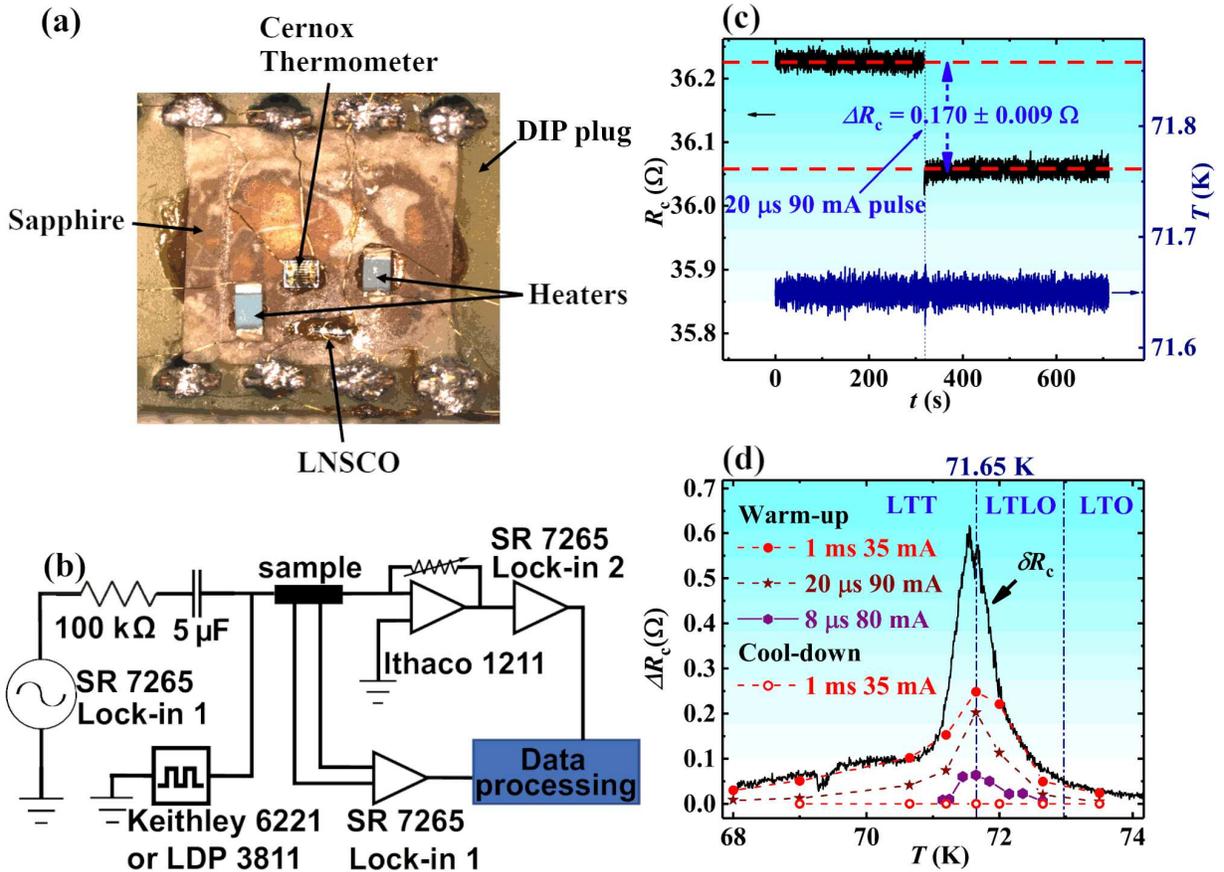}
    \caption{(a) The two heaters (surface-mount metal-film resistors), a Cernox thermometer, and the sample are mounted on the same sapphire platform on top of a 16-pin DIP plug made of G-10, with a Si diode mounted underneath it (not shown).  The heaters, connected in parallel, are placed on the opposite sides of the sample to try to avoid any thermal gradients across the sample.  25-$\mu$m gold wires are used to make electrical contacts to the DIP plug.  (b)  The SR7265 lock-in amplifier 1 along with a 100 k$\Omega$ resistor provides a small ac current $I=10~\mu$A, which is measured by the SR7265 lock-in 2 via Ithaco 1211 current preamplifier. SR7265 lock-in 1 measures the ac voltage, and Keithley 6221 or LDP-3811 current sources generate dc current pulses (see Supplementary Fig.~1 for more details).  (c)  The resistance drop ${\Delta}R_\mathrm{c}$ after applying a 20-$\mu$s, 90-mA pulse at 71.65~K.  The vertical, thin dotted blue line shows the moment when the pulse is applied.  $T$ is shown as measured by the Cernox (blue trace; right y-axis).  (d) ${\Delta}R_\mathrm{c}$ obtained after applying current pulses of various $I_\mathrm{p}$ and $\tau$, as shown, at different $T$.  Solid (open) symbols: ${\Delta}R_\mathrm{c}$ after a warm-up (cool-down) protocol.  No resistance drops are observed after cooling. Black curve: $\delta R_\mathrm{c}$, the difference between the warming and cooling branches of the hysteresis loop (Fig.~\ref{fig:RT}).  For all pulses, the maximum ${\Delta}R_\mathrm{c}$ is observed at $T\approx T_{\mathrm{CO}}\lesssim T_{\textrm{LTT}}$ after the warm-up protocol.} 
    \label{setup}
\end{figure*}

The single crystal of \lnsco\ was grown using the traveling-solvent floating-zone technique.  We measure $R_\mathrm{c}$ on a bar-shaped sample with dimensions 0.24~mm$\times$0.41~mm$\times$1.46~mm ($a\times b\times c$), between the voltage contacts placed at a distance $\approx0.25$~mm.  The contacts are made by attaching gold leads ($\approx25~\mu$m thick) using the DuPont 6838 Ag-paste, followed by a heat treatment at 450~$^{\circ}$C in the flow of oxygen for 30 min. The resulting contact resistances are less than 0.5~$\Omega$ at both room temperature and $\sim70$~K.  The sample, one Cernox thermometer (CX-1070-BG-HT, serial X92666), and two surface-mount metal-film resistors (as heaters) are placed on the same sapphire platform on top of the 16-pin DIP plug made of G-10 [Fig.~\ref{setup}(a)]. 
The precise $T$ control at the sample is achieved by a Lake Shore 336 temperature controller using the heaters and the Cernox thermometer; temperature reading from the Cernox is used as the nominal sample $T$.  A Si diode is fixed beneath the 16-pin DIP plug as a secondary thermometer to monitor $T$ stability.  For better $T$ control, a probe thermometer and a vaporizer temperature are also monitored during the measurement.  The probe thermometer, also a Cernox, is quite far from the sample and is controlled by a probe heater. The probe thermometer is used to sweep or maintain temperature coarsely. The vaporizer temperature is monitored to keep the flow of liquid helium constant during the measurement; this is obtained by fine-tuning the needle-valve opening of the variable-temperature insert and pumping the sample space using a roughing pump. 

$R_\mathrm{c}$ is measured using either a Keithley 6221 current source and 2182A nanovoltmeter in delta mode or SR 7265 lock-in amplifiers using a standard four-probe ac method ($\sim157$~Hz).  Relatively longer pulses ($\tau\geq1$~ms) are generated using the Keithley instruments, controlled with a home-made LabVIEW program, or using the LDP-3811 precision current source.  For shorter pulses ($\tau\geq2~\mu$s), LDP-3811 is used together with the lock-ins [Fig.~\ref{setup}(b)].  The results did not depend on the choice of instrumentation.  The output of the LDP-3811 actually consists of two pulses separated by 100~ns; hereafter, we refer to this sequence as a ``pulse'' (e.g., a 20-$\mu$s, 20-mA  ``pulse'' consists of two 10-$\mu$s, 20-mA pulses, the second one starting 100 ns after the first pulse ends).  Current pulses are applied after a measurement $T$ is reached by following either the ``warm-up'' or the ``cool-down'' protocols.

In the warm-up protocol, the sample is first cycled across the hysteresis by warming up to 90~K and cooling back down to 40~K, followed by warming to a temperature slightly lower than the intended temperature using the probe heater at a rate of 1~K/min.  Then, using the heaters near the sample, the measurement $T$ is reached at a slower rate, typically 0.1~K/min, to avoid overshooting of $T$.  In the cool-down protocol, the measurement $T$ is approached from above: first, the sample is cycled across the hysteresis by cooling down to 40~K and warming up to 90~K using probe heater at a rate of 1~K/min, then the probe heater is used at a rate of 1~K/min to reach a temperature slightly higher than the intended temperature and, finally, metal-film resistors are used to reach the measurement $T$ at a slower rate, typically 0.1~K/min, without an overshoot.  Hereafter, unless stated otherwise, the pulses are applied after the warm-up protocol. 

Figure~\ref{setup}(c) shows a representative effect of a single current pulse on $R_\mathrm{c}$.  The pulse induces switching to a stable, lower resistance state, with ${\Delta}R_\mathrm{c}$ defined as the drop in $R_\mathrm{c}$ after the pulse.  Similar measurements are performed with different $I_\mathrm{p}$ and $\tau$ at various $T$, with each measurement carried out after either a warm-up or a cool-down protocol.  ${\Delta}R_\mathrm{c}$ has a maximum at $T\approx T_{\textrm{CO}}\lesssim T_{\textrm{LTT}}$ [Fig.~\ref{setup}(d)], where the difference $\delta R_c$ between the warming and cooling branches of the main hysteresis loop in Fig.~\ref{fig:RT} is also maximum.  Notably, the resistance drops are observed only after the warm-up protocol, i.e. when the measurement $T$ is approached from a CO phase, consistent with the asymmetry observed in the prior study\cite{Baity2018} of CO dynamics in \lnsco, and suggesting that current pulses induce switching in a CO system into different metastable states.  The question is whether such pulse-induced metastable states are a)  caused by nonthermal effects of the current, or b) they result from the Joule heating of the CO system during the pulse and its subsequent cooling to the bath $T$.  In the latter scenario, ${\Delta}R_\mathrm{c}$ would be observed simply because the system follows the hysteretic $R_\mathrm{c}(T)$ behavior.  

\begin{figure}[!hbt]
    \centering
           \includegraphics[width= 0.45 \textwidth]{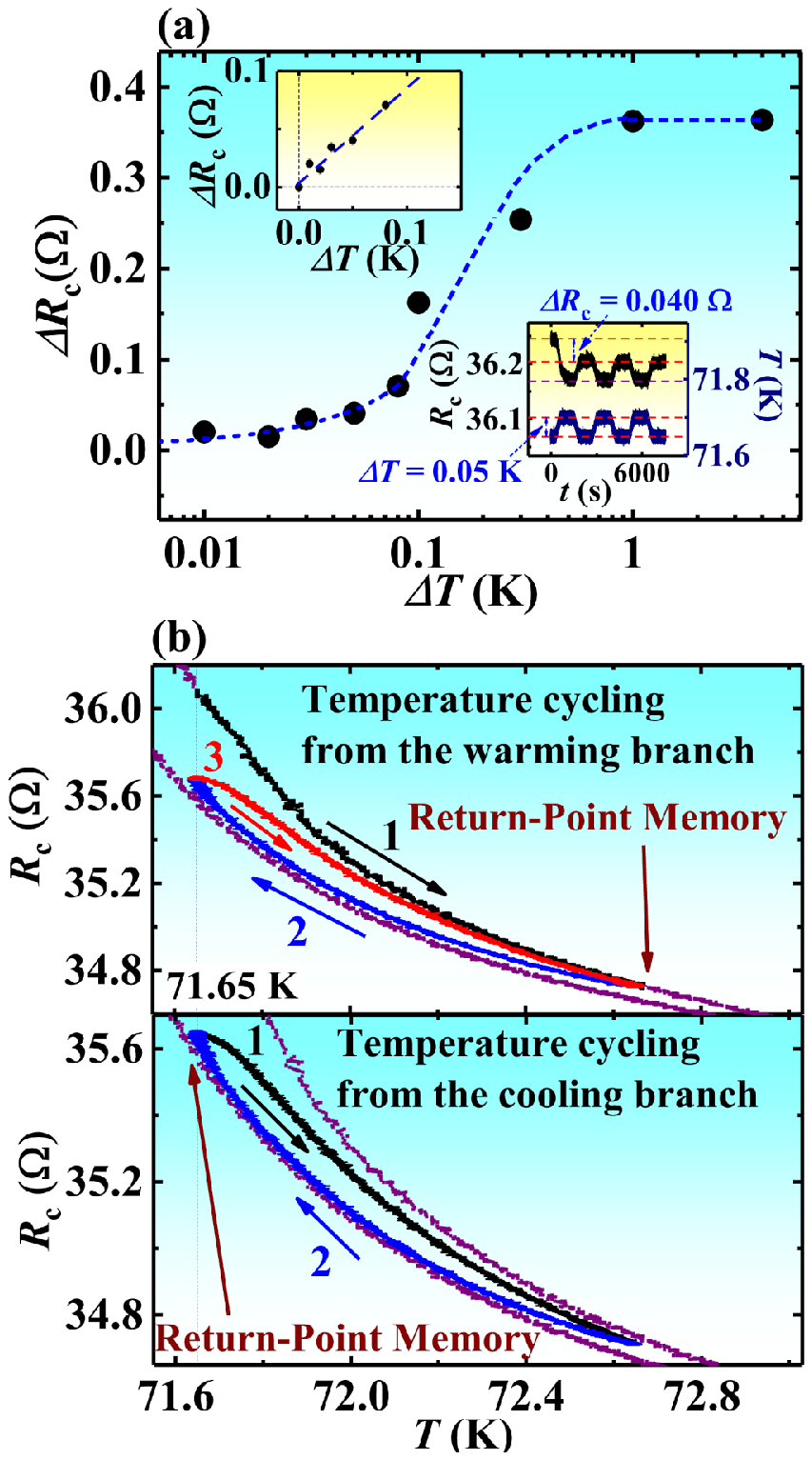}
    \caption{(a) ${\Delta}R_\mathrm{c}$ vs $\Delta T$ obtained at $T=71.65$~K after applying a heat pulse on the sample, i.e. after increasing and then decreasing $T$ by a fixed $\Delta T$ (lower-right inset and Supplementary Fig.~2) using heaters.  The blue dashed line guides the eye.  Top-left inset: ${\Delta}R_\mathrm{c}\propto\Delta T$, with the slope $\sim1~\Omega$/K, for small $\Delta T$; ${\Delta}R_\mathrm{c}=0$ for ${\Delta}T=0$.  (b) The behavior in (a) is consistent with the thermal hysteresis shown here.  The arrows and numbers describe the direction and the order of $T$ sweeps starting from the same $T=71.65$~K.  $R_\mathrm{c}(T)$ exhibits return-point memory observed on cycling the temperature up and down by $\Delta T$ from the warming (cooling) branch shown on the top (bottom), i.e. $R_\mathrm{c}$ does not change if the same $\Delta T$ cycle is repeated.    }     
    \label{results2}
\end{figure}

To explore the possibility of heating, we apply a ``heat pulse''', i.e. we increase and then decrease $T$ by a fixed ${\Delta}T$ (Fig.~\ref{results2}(a), lower-right inset; also Supplementary Fig.~2).  We find that
$\Delta R_\mathrm{c}$ depends only on $\Delta T$, and it does not depend on the number of subsequent heat pulses with the same $\Delta T$.  $\Delta R_\mathrm{c}$ increases with $\Delta T$ and saturates for $\Delta T\gtrsim $1~K [Fig.~\ref{results2}(a)].  Importantly, $\Delta R_{\mathrm{c}}\propto\Delta T$ at low $\Delta T$ (Fig.~\ref{results2}(a), upper-left inset), indicating that $\Delta R_\mathrm{c}$ vanishes as $\Delta T\rightarrow 0$.  These results are indeed consistent with the presence of a thermal hysteresis in $R_\mathrm{c}(T)$.  For example, if a heat pulse is applied at $T=71.65$~K after a warm-up protocol (Fig.~\ref{results2}(b), top), $R_\mathrm{c}$ follows the main warming branch of the hysteresis (black trace, arrow marked 1), followed by cooling along a subloop marked by (blue) arrow 2, resulting in a lower $R_\mathrm{c}$ once back at the initial $T=71.65$~K.  Any subsequent heat pulse with the same $\Delta T$ will keep $R_\mathrm{c}$ on the same subloop (blue-red, arrows 2 and 3), as the system exhibits return-point memory.  The return-point memory was found also in the magnetoresistance hysteresis in the same material.\cite{Baity2018}  Figure~\ref{results2}(b), bottom shows the subloop (black-blue, arrows 1 and 2) obtained when the heat pulse is applied at $T=71.65$~K after a cool-down protocol.  In that case, $\Delta R_\mathrm{c}=0$ is expected after a heat pulse, as observed.

\begin{figure}[!hbt]
    \centering
         \includegraphics[width= 0.45 \textwidth]{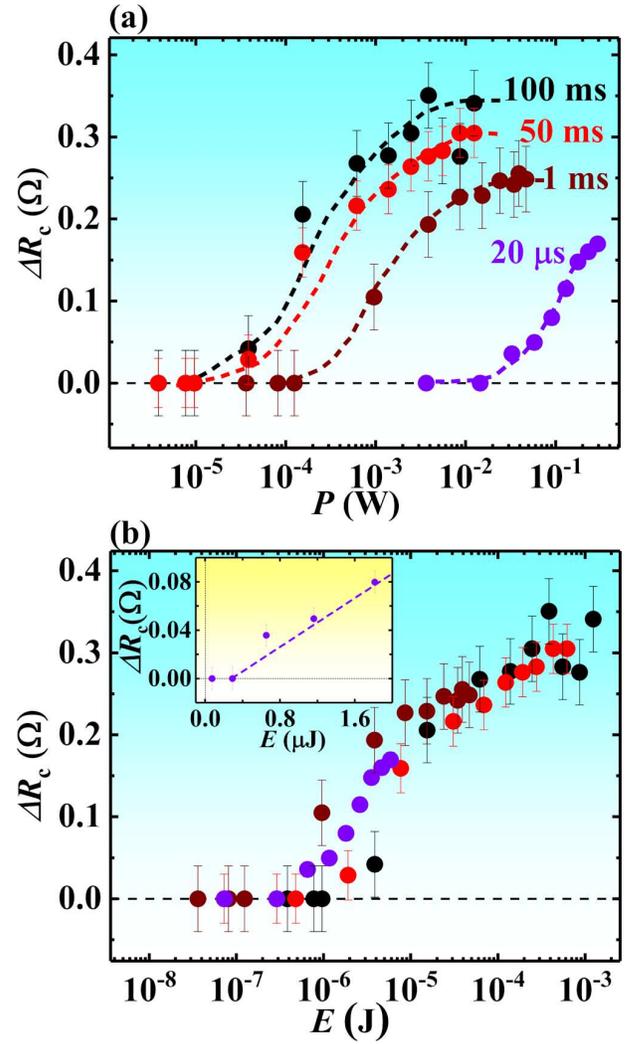}
    \caption{(a) $\Delta R_\mathrm{c}$ obtained after applying a current pulse plotted vs power $P$ for different pulse duration $\tau$, as shown.  (b) The data in (a) plotted as a function of energy $E$ injected into the system.  Inset: ${\Delta}R_\mathrm{c}$ vs $E$ for $\tau=20~\mu$s on a linear scale near $E=0$.}    
    \label{results3}
\end{figure}

The effects of electric pulses are different from those of heat pulses.  First, we examine the dependence of $\Delta R_\mathrm{c}$ on the power applied to the sample during a single pulse, $P \approx {I_\mathrm{p}}{^2}R{_\mathrm{c}}$, and on the energy injected into the system, $E\approx P\tau$, where $ I\ll I_\mathrm{p}$ and $R_\mathrm{c}$ is the resistance state before applying the pulse.  ($P$ and $E$ are thus calculated for the fraction of the sample volume where $R_\mathrm{c}$ is measured.  The dependence of $\Delta R_\mathrm{c}$ on $I_\mathrm{p}$ is shown in Supplementary Fig.~3.)  It is obvious that, for each $\tau$, there is a  threshold power below which no resistance drop is observed, followed by an increase in $\Delta R_\mathrm{c}$, and then a tendency towards saturation at the highest $P$ [Fig.~\ref{results3}(a)].  Similar behavior is observed as a function of injected energy  [Fig.~\ref{results3}(b)], with an important difference that the data for all different $\tau$ and $I_\mathrm{p}$ scale with $E$ and exhibit 
the same threshold energy $\sim (4-10)\times 10^{-7}$~J.  The scaling of ${\Delta}R_\mathrm{c}$ with $E$ indicates poor thermal coupling of the electronic system to the environment during $\tau$, such that the system cannot reach thermal equilibrium with the bath during the application of a pulse. Although some heating might be expected, especially for high values of $E$, the existence of a threshold, absent in the case of heat pulses (Fig.~\ref{results2}(a) inset), suggests that nonthermal processes dominate at low $E$.  

\begin{figure*}[!ht]
    \centering
           \includegraphics[width= 0.90 \textwidth]{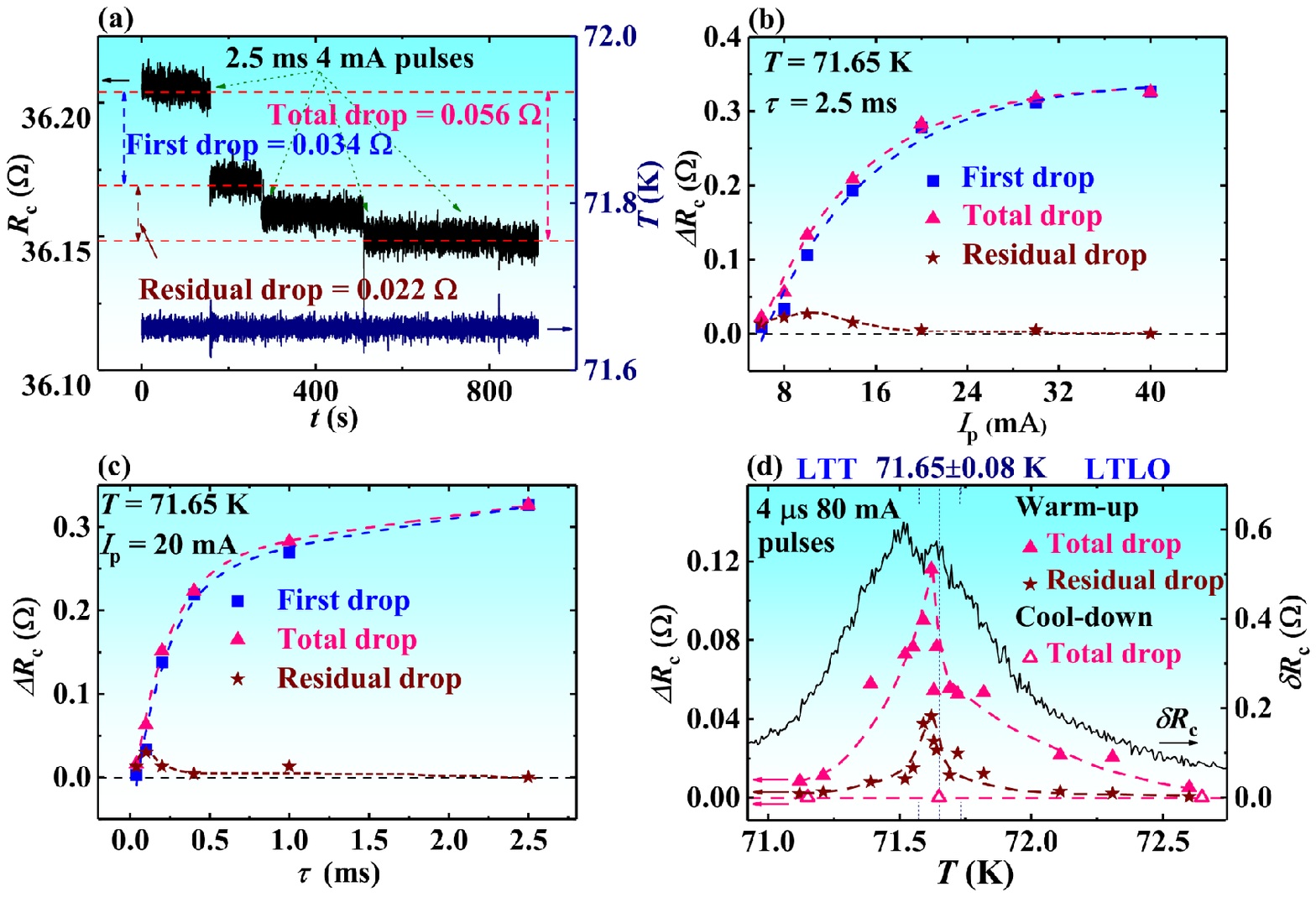}
    \caption{(a) Multiple identical pulses applied at $T=71.65$~K after a warm-up protocol, generating multiple drops in $R_\mathrm{c}$, in contrast to the heat pulses in Fig.~\ref{results2}(a).  The first $\Delta R_\mathrm{c}$ drop is marked as the ``first drop'', and the rest are marked as the ``residual drop''. On applying the pulses many times, $R_\mathrm{c}$ does not change anymore and the sum of all the resistance drops is marked as the ``total drop''.  The first, residual, and total drops at $T=71.65$~K vs (b)  $I_\mathrm{p}$ for $\tau=2.5$~ms, and (c) $\tau$ for a fixed $I_\mathrm{p}=20$ ~mA.  (d) The drops obtained after applying 8-$\mu$s, 80-mA current pulses multiple times at various temperatures.  All the drops show a peak at $T\approx T_{\mathrm{CO}}\lesssim T_{\textrm{LTT}}=71.65$~K.} 
        \label{results4}
\end{figure*}

Next, we apply electric pulses multiple times.  Figure~\ref{results4}(a) shows the data obtained with 2.5-ms, 4-mA pulses applied four times following the initial warm-up protocol 
(i.e. the warm-up protocol was performed only before the first pulse).  The first three pulses cause observable drops in $R_\mathrm{c}$, and further pulse application does not result in any change of $R_\mathrm{c}$.  This behavior is significantly different from the effect of heat pulses.  For example, the ``first drop'' in $R_\mathrm{c}$, produced by the first pulse, is about $0.034~\Omega$; if the entire effect of the pulse was Joule heating, this would correspond to an increase in temperature by $\Delta T\sim0.04$~K [Fig.~\ref{results2}(a)] and there should be no change in $R_\mathrm{c}$ after additional pulses are applied, in contrast to our findings.  This provides additional evidence that an electric pulse in this case causes predominantly nonthermal effects.

To explore the conditions necessary to overcome the thermal regime, we determine both the first drops and the ``residual drops'', i.e. $\Delta R_\mathrm{c}$ produced by all subsequent pulses, as a function of $I_\textrm{p}$ for a fixed $\tau$ [Fig.~\ref{results4}(b)] and as a function of $\tau$ for a fixed $I_\textrm{p}$ [Fig.~\ref{results4}(c)].  The ``total drop'' is defined as the sum of the first and residual drops.  In both cases, we see similar behavior: the first and total drops increase rapidly with $I_\textrm{p}$ (and $\tau$), followed by a much weaker dependence or quasi-saturation at high $I_\textrm{p}$ (and long $\tau$).  However, the residual drop exhibits a different trend, showing an enhancement, i.e. a broad peak, at $I_{\mathrm{p}}\approx$~3-7~mA [Fig.~\ref{results4}(b)] or $\tau\approx100~\mu$s [Fig.~\ref{results4}(c)], before vanishing at higher values of $I_\mathrm{p}$ and $\tau$.  This indicates that, although Joule heating might dominate at large $I_\textrm{p}$ and long $\tau$, for small perturbations the situation is different: here pulsed current injection allows access to nonthermally-induced resistive states.  The similarities in the dependence of various drops $\Delta R_\mathrm{c}$ on $I_\mathrm{p}$ and on $\tau$ signify that it is again the energy injected into the system that plays a major role, i.e. that determines the size of the resistance drops.  Indeed, it is only for longer pulses ($\tau\gtrsim 1$~ms) that $\Delta R_\mathrm{c}$, for a fixed $E$, starts to depend also on $\tau$ (Supplementary Fig.~4), indicating that the system is no longer thermally isolated from the bath.  

Finally, by using multiple-pulse current injection in the regime where nonthermal effects dominate, we probe the current-induced resistive metastable states as a function of $T$.  We find that the first, residual, and total drops all have a sharp peak at $T\approx T_{\mathrm{CO}}\lesssim T_{\textrm{LTT}}=71.65$~K, and that the drops are observed only after a warm-up protocol [Fig.~\ref{results4}(d)].  The asymmetry of the observed nonequilibrium states is 
analogous to that found\cite{Baity2018} by studying sharp resistance drops or avalanches resulting from a change of the applied $H$.  In addition, the avalanches were observed only above a threshold $H\sim 2$~T, which was thus identified as the minimum depinning field for the CO domains.  (The stripe correlation length in \lnsco\ is $\sim 11$~nm.\cite{Wilkins2011})  Our study of the pulsed current injection has revealed that there is indeed a threshold energy that needs to be injected into the system to induce switching into another resistive metastable state.  In contrast to the magnetoresistance study that showed\cite{Baity2018} two peaks in the avalanche occurrence, a stronger one in the LTLO phase and a weaker one in the LTT phase, tentatively attributed to the onset of precursor nematic order and CO, respectively, we find only one peak, sharp and somewhat asymmetric, such that $\Delta R_\mathrm{c}$ is more pronounced on the LTT side of the transition.  In addition, there is no evidence of metastable states in the LTO phase.  All the results are consistent with pinning of the fluctuating CO, with fluctuations becoming weaker away from the transition, in agreement with general expectations.

We have established that pulsed current injection is a viable and effective method for probing the CO domain dynamics in cuprates.  Previous attempts to detect collective stripe motion in cuprates\cite{Lavrov2003} and nickelates\cite{Huecker2007,Pautrat2007} using high electric fields, i.e. by measuring current-voltage characteristics, found only nonlinear transport effects that could be attributed to Joule heating.  The effects of current pulses were either estimated\cite{Lavrov2003} or explored\cite{Huecker2007} for very long pulse duration ($\tau\approx200$~ms).  However, heating effects are generally not easy to estimate because they depend on a variety of factors in a given experimental set-up, including the sample substrate and the cooling power of the cryostat.  In contrast, we have demonstrated a systematic way to investigate the effects of pulsed current injection and experimentally identify the regime in which nonthermal effects dominate.  This has allowed us to detect signatures of the fluctuating CO in \lnsco, thus paving the way for similar studies in other materials.

\section*{Supplementary Material}
See supplementary material for more details about the measurement set-up, additional ${\Delta}R_\mathrm{c}$ vs $\Delta T$ data, the dependence of $\Delta R_\mathrm{c}$ on $I_\mathrm{p}$, and the results from multiple-pulse current injection while keeping the energy injected into the system fixed.

\begin{acknowledgments}
We are grateful to D. Smirnov for help with the instrumentation.  This work was supported by NSF Grant No.~DMR-1707785, and the National High Magnetic Field Laboratory through the NSF Cooperative Agreement No. DMR-1644779, DMR-1644779 via User Collaboration Grants Program 5206, and the State of Florida.
\end{acknowledgments}

\section*{Author contributions}
Single crystals were grown and prepared by T. S.; B. K. P., Y. W., and J. J. performed the measurements and analyzed the data; B. K. P., Y. W. and D. P. wrote the manuscript, with input from all authors; D. P. supervised the project. 

\section*{Data availability}
The data that support the findings of this study are available from the corresponding author upon reasonable request.

\clearpage

\renewcommand{\figurename}{{SUPPLEMENTARY FIG.}}

\setcounter{figure}{0}

\onecolumngrid{

\begin{center}
\large\textbf{Supplementary Material  \\
Charge order dynamics in underdoped La$\mathbf{_{1.6-\textit{x}}}$Nd$\mathbf{_{0.4}}$Sr$\mathbf{{_\textit{x}}}$CuO$\mathbf{_{4}}$ revealed by electric pulses}
\vspace{12pt}

\normalsize

Bal~K.~Pokharel,$^{1,2}$ Yuxin~Wang,$^{1,2}$ J.~Jaroszynski,$^1$  T. Sasagawa,$^3$ and Dragana Popovi\'c$^{1,2}$
\vspace{6pt}

\small
$^1$\textit{National High Magnetic Field Laboratory, Tallahassee, Florida 32310, USA}\\
$^2$\textit{Department of Physics, Florida State University, Tallahassee, Florida 32306, USA}\\
$^3$\textit{Materials and Structures Laboratory, Tokyo Institute of Technology, Kanagawa 226-8503, Japan}

\end{center}
}


\setlength{\textfloatsep}{5pt plus 1.0pt minus 2.0pt}

\begin{figure*}[h]
    \centering
     \includegraphics[width= 0.90 \textwidth]{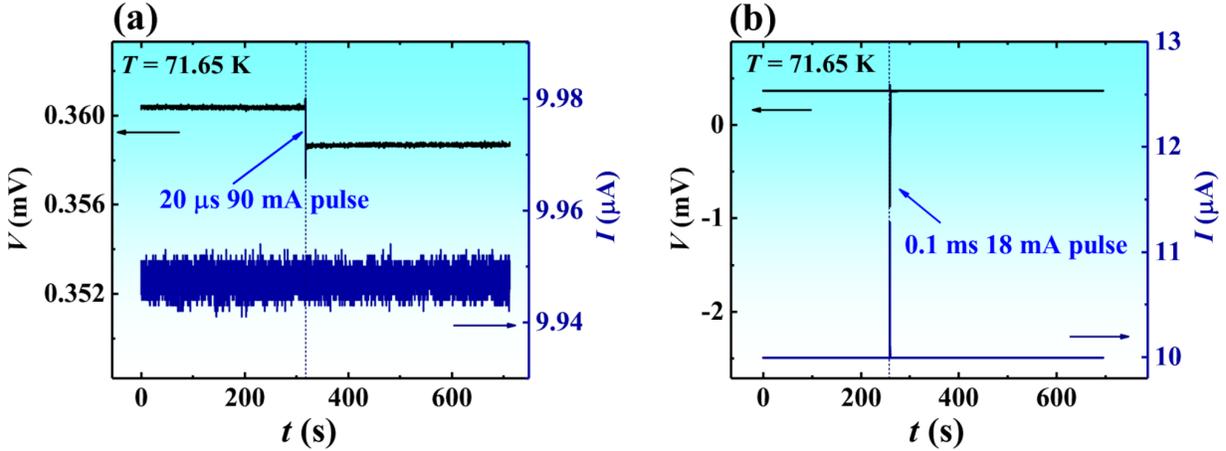}
    \caption{ Representative measurements of the ac voltage $V$ and the ac current $I$ vs time $t$ for current pulses with (a) $\tau=20~\mu$s, $I_\mathrm{p}=90$~mA, and (b) $\tau=0.1$~ms, $I_\mathrm{p}=18$~mA.  The sensitivity of Ithaco 1211 current preamplifier [Fig.~2(b)] was $10^{-3}$~A/V, with the corresponding input resistance of $0.5~\Omega$.  The dc current pulse produces a spike in both $V$ and $I$ readings, but $I$ is the same before and after the pulse.  The drop in $V$ after the pulse thus reflects a drop in $R_\mathrm{c}$.  It was verified that the same results, both before and after the pulse, were obtained when Ithaco 1211 current preamplifier was removed from the circuit.  In any case, $V$ and $I$ readings during the pulse were not used in any of our analysis.}
    \label{I-and-V}
\end{figure*} 

\begin{figure*}[h]
    \centering
     \includegraphics[width= 0.98 \textwidth]{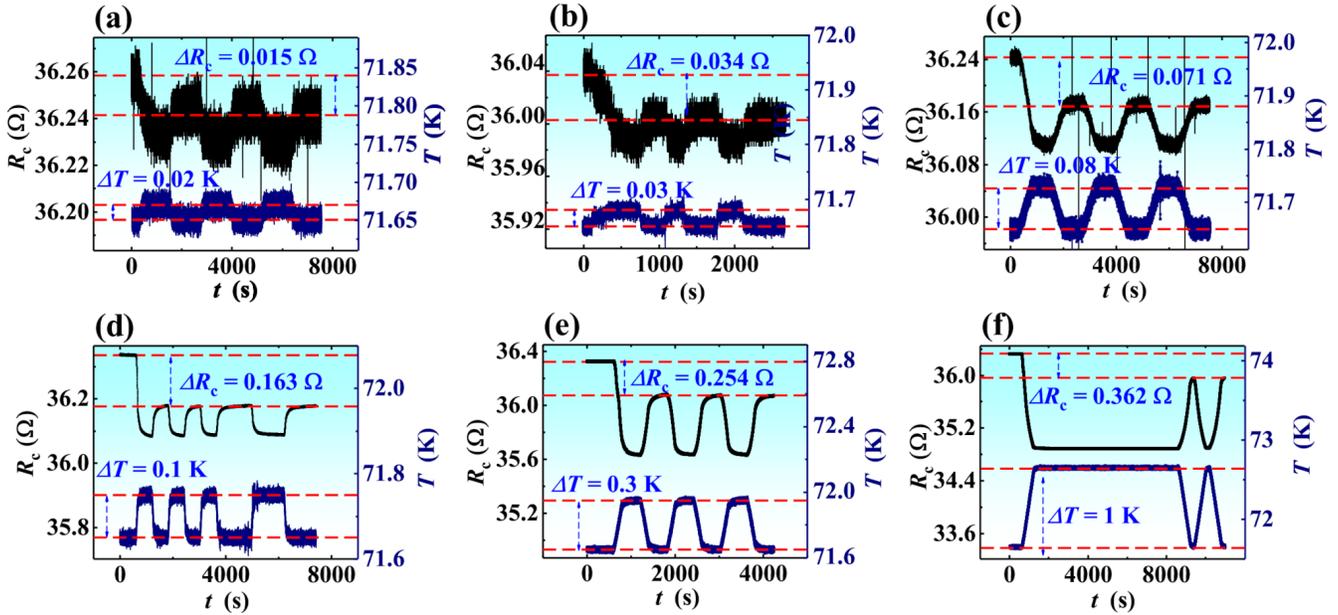}
    \caption{ The effect seen in $R_\textrm{c}$ on increasing the temperature by various amounts $\Delta T$, as shown, starting from $T=71.65$~K.  Increasing and decreasing the temperature many times does not change the resistance states reached after the first increase and decrease. }
    \label{supp-DeltaT}
\end{figure*} 

\begin{figure*}[h]
    \centering
         \includegraphics[width=17.0cm,clip=]{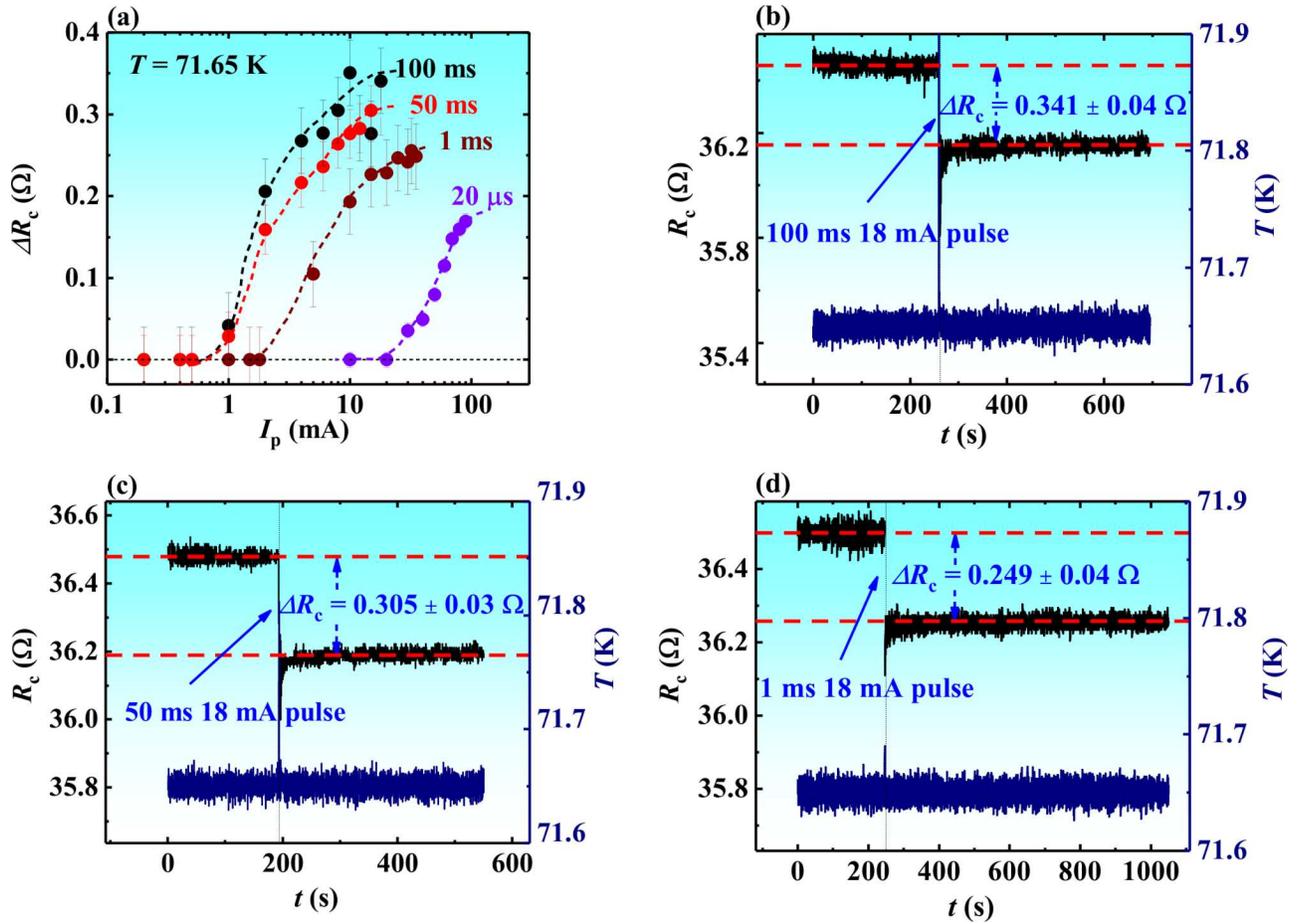}
    \caption{(a) ${\Delta}R_\mathrm{c}$ vs. pulse current magnitude, $I_\textrm{p}$, for different duration $\tau$ of 100 ms, 50 ms, 1 ms, and 20 $\mu$s, at $T=71.65$~K.  A rapid rise of ${\Delta}R_\mathrm{c}$ with $I_\mathrm{p}$ is observed, with a threshold pulse current magnitude that depends on $\tau$.  (b), (c) and (d) show the effect of single pulses with the same magnitude ($I_\mathrm{p}=18$~mA) but different duration $\tau=$~100~ms, 50~ms, 1~ms, respectively.} 
    \label{supp3}
\end{figure*}

\begin{figure*}
    \centering
            \includegraphics[width= 0.85 \textwidth]{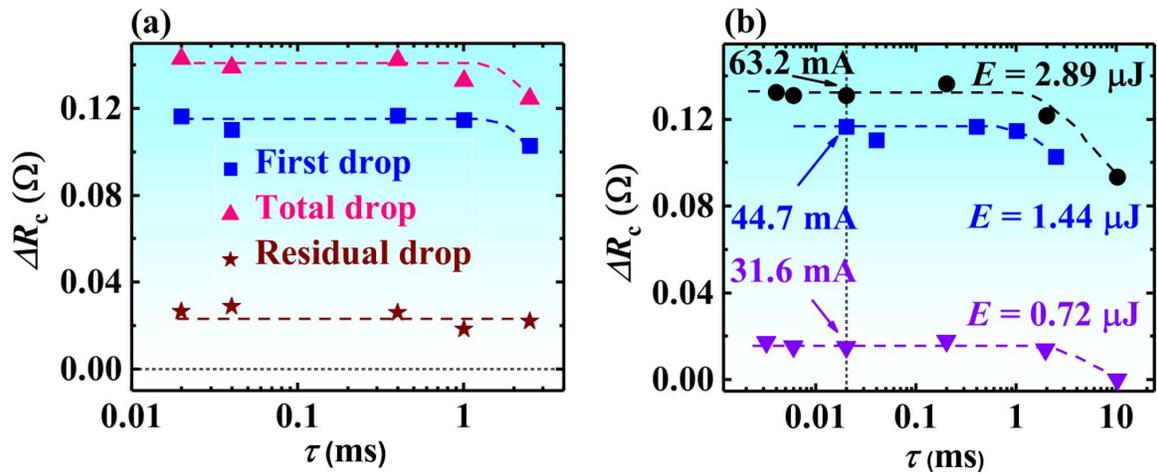}
    \caption{The pulse duration and magnitude of the current pulse are varied to keep the energy injected into the system fixed. For a fixed energy, multiple pulses are applied to determine the first, residual, and total drops.  (a) The results for $E=1.44~\mu$J are summarized as a function of $\tau$.  Except for the longest pulses ($\tau\gtrsim 1$~ms), all drops remain independent of $\tau$.  (b) The first drops are shown for three more energies, 2.89~$\mu$J, 1.44 ~$\mu$J, and 0.72~$\mu$J.  All energies are calculated for a single pulse. } 
    \label{supp2}
\end{figure*}

\end{document}